\documentclass[11pt,a4paper]{article}

\usepackage[margin=1in]{geometry}
\usepackage[T1]{fontenc}
\usepackage[utf8]{inputenc}
\usepackage{lmodern}
\usepackage{microtype}
\usepackage{graphicx}
\usepackage{booktabs}
\usepackage{amsmath,amssymb}
\usepackage{caption}
\usepackage[hidelinks,colorlinks=true,linkcolor=blue!50!black,
            citecolor=blue!50!black,urlcolor=blue!50!black]{hyperref}
\usepackage{xcolor}
\usepackage{tikz}
\usetikzlibrary{shapes,arrows.meta,positioning,fit,backgrounds,calc,decorations.pathreplacing}
\usepackage{natbib}

\title{\textbf{The Conversations Beneath the Code:\\
Triadic Data for Long-Horizon Software Engineering Agents}}

\author{Yelin Kim \\
        \normalsize Independent Researcher \\
        \normalsize \texttt{yelinkim@umich.edu}}

\date{May 2026}

\begin{document}
\maketitle

\begin{abstract}
\noindent
Frontier software engineering agents have saturated short-horizon benchmarks
while regressing on the work that constitutes senior engineering: long-horizon,
multi-engineer, ambiguous-specification deliverables. This paper takes a
position on what training data is needed to close the gap. The substrate for
the next generation of SWE agents is neither larger GitHub scrapes nor more
solo-agent trajectories nor---sufficient by itself---open human--AI dialogue
logs. It is \emph{triadic data}: synchronized capture of the human--human
conversations where engineering context is formed, the human--AI sessions
where that context is partially consumed, and the multi-week cross-functional
work that surrounds both. We argue that the canonical instantiation of triadic
data is two complementary products: long-horizon expert trajectories captured
under stimulated-recall protocols, and \emph{simulated cross-functional
companies}---instrumented teams of senior engineers, product managers,
designers, and data scientists working through ambiguous deliverables on
shared infrastructure. We further specify a four-tier evidence framework
through which any such corpus---triadic or otherwise---must justify its
quality to a fine-tuning researcher: mechanical verification, statistical
corpus characterization, probe experiments, and pre-registered blind
evaluation. We argue that this data is capturable in 12--18 months with
methods already mature in adjacent fields, that it is the empirical key to
four open questions in agent training, and that the field's near-term
research agenda should include it explicitly.

\end{abstract}

\section{The Position}

The state of frontier software engineering agents in mid-2026 contains a
contradiction that the data agenda has not yet caught up to.

On one side, short-horizon benchmarks have saturated. SWE-bench Verified moved
from 13.8\% (Devin, March 2024) to 82.0\% (Claude Sonnet 4.5, September 2025).
Terminal-Bench 2.0 top scores sit at 82.0\%~\citep{terminalBench2026}.
Aider's polyglot leaderboard places GPT-5 at 88.0\% across six languages.
Whatever signal these benchmarks were carrying, frontier models have absorbed
it.

On the other side, long-horizon benchmarks are not saturating. METR's Time
Horizon measurement program~\citep{metrTH2025} places frontier models at
roughly 100\% success on tasks taking humans under four minutes and under
10\% on tasks taking over four hours. SWE-EVO~\citep{sweEVO2025}, designed
around 48 long-horizon repository-evolution tasks, reports the best frontier
model at 25.0\%---against 72.8\% on standard SWE-bench Verified for the same
model class. SWE-Lancer~\citep{sweLancer2025}, evaluating frontier models on
real freelance work, reports models ``unable to solve the majority of tasks.''
The gap between short and long horizons is not closing. It is being revealed.

The dominant explanation in the literature is that this is a
\emph{context-handling} gap, not a reasoning gap. Anthropic's harness-design
literature~\citep{anthropicHarness2025,anthropicManaged2026} centers context
engineering as the runtime lever, with proposed fixes (context resets,
sub-agent decomposition, event virtualization) acting as workarounds for the
absence of a principled training signal. The APEX-SWE
benchmark~\citep{apexSWE2026} identifies ``epistemic discipline''---the
ability to distinguish what was assumed from what was verified---as the
dominant skill gap, with frontier models clustering in the 30--40\% pass@1
range on Integration and Observability tasks. METR's failure-mode analysis
shows that reasoning quality on subproblems extracted from long-horizon
failures matches the corresponding short-horizon benchmarks; what degrades is
state handling across time, not reasoning per token.

If reasoning is solved on short horizons and the bottleneck on long horizons
is context, then the next generation of SWE agents will be built on training
data that captures how senior engineers handle context. \textbf{That data does
not exist at scale in any public corpus.} It is not in GitHub commits, which
capture artifacts and not deliberation. It is not in agent traces, which
capture what agents did with context but not how the context was formed. It
is not in human--AI dialogue logs alone---though those are valuable and
underexploited~\citep{wangHAICoding2026}---because the substantive engineering
deliberation usually happens before the developer sits down with the AI.

The data lives in human--human conversations: design reviews, on-call
hand-offs, pair programming, architecture debates, postmortems, the ambient
triage that happens in Slack, over whiteboards, and across tickets. The data
also lives in the cross-functional friction between engineers, product
managers, designers, and data scientists working through ambiguous
deliverables over weeks. And it lives in the moments where these conversations
meet AI tools---where context formed in human deliberation is partially
surfaced in a prompt, partially missed, partially recovered through
correction.

We call this \emph{triadic data}: human--human--AI rather than human--AI. We
argue that capturing it at scale, primarily through simulated cross-functional
companies and instrumented long-horizon expert sessions, is the most leveraged
training-data investment available to frontier SWE labs over the next 12--18
months.

The remainder of this paper develops the position in seven movements.
Section~\ref{sec:insufficient} establishes why existing data approaches are
insufficient, including the recent and important call by~\citet{wangHAICoding2026}
for open human--AI dialogue data, which we endorse as necessary but argue is
insufficient. Section~\ref{sec:triadic} develops the triadic frame and its
three configurations. Section~\ref{sec:products} specifies the two complementary
data products and their methodologies. Section~\ref{sec:quality} introduces
a four-tier evidence framework---ported from a prior data-quality memo by the
author---through which a triadic corpus (or any post-training corpus) must
justify itself to a fine-tuning researcher. Section~\ref{sec:directions}
outlines four research directions the substrate enables.
Section~\ref{sec:deprioritize} names what this approach deprioritizes and
what would invalidate the position. Section~\ref{sec:future} sketches future
directions, including the perception-native and visual-SWE territory where
triadic methodology has natural extensions.

\section{Why Existing Data Approaches Are Insufficient}\label{sec:insufficient}

The training-data agenda for SWE agents has consolidated around four
approaches over the past three years. Each has produced real gains. None is
sufficient for the long-horizon regime.

\subsection{GitHub-scale scraping has hit diminishing returns}

Public code corpora capture artifacts, not deliberation.
SWE-bench+~\citep{swebenchPlus2024} audited successful patches across the
standard SWE-bench resolution set and found 32.67\% had solution leakage in
the issue text and 31.08\% had tests too weak to verify correctness; filtered
resolution rates dropped from 12.47\% to 3.97\%---a threefold inflation.
Independent quality analysis from GitClear~\citep{gitclear2024} shows code
churn doubled in the post-AI period, consistent with models trained on
pattern-rich tutorial code learning to produce fluent but architecturally
fragile output. The signal is not absent, but the marginal trajectory is
unfavorable.

\subsection{Solo-agent self-trajectories suffer cold-start and bias problems}

Reinforcement learning from agent-generated trajectories in environments with
verifiable rewards has driven recent gains, but the mechanism amplifies what
models already do rather than introducing what experts do.
\citet{yueRLVR2025} argue that reinforcement learning with verifiable rewards
focuses existing capabilities rather than creating new ones. Synthetic data
has diminishing returns. Without an exogenous expert signal, the regime risks
the long-horizon equivalent of model collapse---agents that are increasingly
fluent in the failure modes of their predecessors.

\subsection{Dyadic human--AI dialogue is necessary but insufficient}

\citet{wangHAICoding2026}, in the position paper ``Humans are Missing from
AI Coding Agent Research,'' correctly observe that ``open conversation data
between humans and AI coding systems remains starkly lacking.'' We endorse
this observation and the call to collect such data. But the dyadic frame---what
the developer typed, what the agent generated, what was accepted or
rejected---captures only what happens after the developer sits down with the
agent. By the time the developer is typing into a coding system, the
substantive engineering deliberation has already happened, mostly with other
engineers, mostly off-record. The dyadic frame sees the prompt; it does not
see the meeting that produced the prompt.

\subsection{The capability gaps converge on the same diagnosis}

Synthesizing the recent gap analyses---long-horizon planning
\citep{metrTH2025,sweEVO2025}, multi-file reasoning under realistic conditions
\citep{swebenchPlus2024}, root-cause debugging, DevOps and infrastructure
\citep{terminalBench2026}, ambiguous requirements \citep{ambigSWE2025},
architectural reasoning, security \citep{perryInsecure2023}, and visual SWE
\citep{swebenchMultimodal2024}---yields a striking pattern. The gaps differ
in surface form but share a common substrate: each is a domain where the
operative knowledge lives in human--human deliberation rather than in code
artifacts. Architectural decisions live in design reviews. Root-cause analysis
lives in pair debugging and postmortems. Ambiguous requirements live in
negotiation between PMs, designers, and engineers. The diagnosis converges:
\textbf{the missing data is the conversation that produces the artifact, not
the artifact itself.}

\subsection{A concrete instance}

The pattern is easier to see through a representative case. Consider an
infrastructure engineer asked to configure Apache Airflow on a new compute
region. The codebase spans multiple regions with different node sizes. Some
configuration is in code; some requires runtime inspection
(\texttt{numactl --hardware}, \texttt{kubectl describe node}); some is
genuinely in nobody's documentation but in the heads of the team that scaled
the cluster the previous Tuesday. The engineer opens an AI agent, types a
prompt, the agent attempts the task.

What governs the success of this interaction is not the quality of the
human--AI dialogue. It is the prior state of the engineer's own context. If
they have just come from a meeting where the cluster-scale change was
discussed, they prompt the agent with that context and the task succeeds. If
they have not, they prompt the agent without it, the agent assumes the
documented configuration is current, and the task fails---possibly silently,
possibly in production. The dyadic frame sees the prompt and the response. It
does not see the meeting, the on-call hand-off, the Slack thread, or the
design review where the change was first agreed.

This is the unit of failure the next generation of training data has to
address.

\section{The Triadic Frame}\label{sec:triadic}

We propose that the unit of analysis for SWE training data should be the
\emph{triad}---two or more humans and at least one AI system---rather than
the dyad. The triad is not a rebranding of multi-modal capture. It is a
claim about where the engineering signal actually lives.

\subsection{Three configurations}

\textbf{Configuration A---Pair-with-AI.} Two engineers work jointly with an
AI agent in real time. One drives the keyboard; the other observes and
intervenes. The AI agent is a third participant---sometimes consulted,
sometimes overridden. The signal value is in the meta-conversation between
the engineers: what one engineer says to the other about what to ask the AI,
when to accept its proposal, when to override.

\textbf{Configuration B---Human--human-then-AI.} Two or more engineers
complete a synchronous deliberation (design review, on-call hand-off,
postmortem) without the AI agent present. One subsequently implements the
agreed plan with AI assistance. The signal is the connection between the
deliberation and the implementation: what context the engineer carried into
the AI session, what was surfaced in the prompt, what was lost.

\textbf{Configuration C---Human--human-around-AI.} Multiple engineers,
possibly across teams, work asynchronously around shared AI agents over days
or weeks. Conversations span Slack, video calls, in-person meetings, code
review, and design docs. The AI agent participates intermittently. This is
the configuration closest to real production engineering at scale, and the
data is multimodal, multi-channel, and longitudinal.

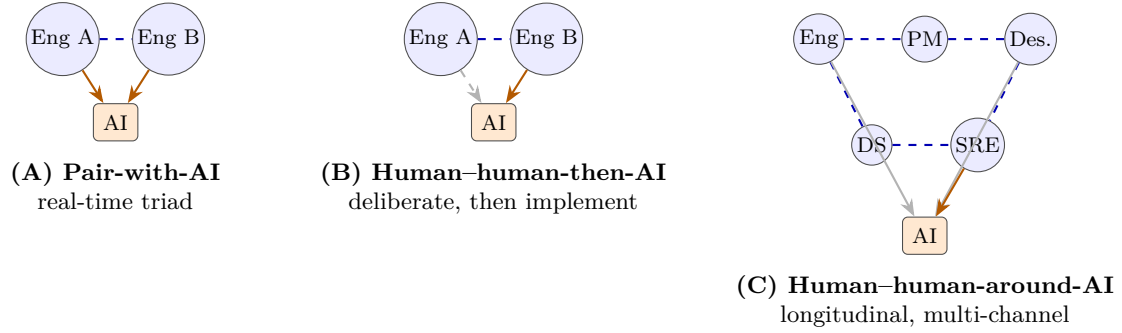
\begin{figure}[t]
  \centering
\begin{tikzpicture}[
  human/.style={circle,draw=black!70,fill=blue!8,minimum size=14pt,inner sep=1pt,font=\scriptsize},
  ai/.style={rectangle,draw=black!70,fill=orange!18,minimum size=14pt,rounded corners=2pt,font=\scriptsize},
  hh/.style={dashed,thick,blue!70!black},
  ha/.style={->,>=Stealth,thick,orange!70!black},
  panel/.style={draw=black!30,rounded corners=4pt,inner sep=8pt,fill=gray!2},
  caplab/.style={font=\footnotesize\bfseries,align=center}
]

\begin{scope}[local bounding box=A]
  \node[human] (Ah1) at (0,0) {Eng A};
  \node[human] (Ah2) at (1.4,0) {Eng B};
  \node[ai]    (Aai) at (0.7,-1.1) {AI};
  \draw[hh] (Ah1) -- (Ah2);
  \draw[ha] (Ah1) -- (Aai);
  \draw[ha] (Ah2) -- (Aai);
\end{scope}
\node[caplab,below=4pt of A] {(A) Pair-with-AI \\ \footnotesize\mdseries real-time triad};

\begin{scope}[shift={(5,0)},local bounding box=B]
  \node[human] (Bh1) at (0,0) {Eng A};
  \node[human] (Bh2) at (1.4,0) {Eng B};
  \node[ai]    (Bai) at (0.7,-1.1) {AI};
  \draw[hh] (Bh1) -- (Bh2);
  \draw[ha,dashed,gray!60] (Bh1) -- (Bai);
  \draw[ha] (Bh2) -- (Bai);
\end{scope}
\node[caplab,below=4pt of B] {(B) Human--human-then-AI \\ \footnotesize\mdseries deliberate, then implement};

\begin{scope}[shift={(10,0)},local bounding box=C]
  \node[human] (Ch1) at (0,0) {Eng};
  \node[human] (Ch2) at (1.4,0) {PM};
  \node[human] (Ch3) at (2.8,0) {Des.};
  \node[human] (Ch4) at (0.7,-1.4) {DS};
  \node[human] (Ch5) at (2.1,-1.4) {SRE};
  \node[ai]    (Cai) at (1.4,-2.6) {AI};
  \draw[hh] (Ch1) -- (Ch2);
  \draw[hh] (Ch2) -- (Ch3);
  \draw[hh] (Ch1) -- (Ch4);
  \draw[hh] (Ch4) -- (Ch5);
  \draw[hh] (Ch3) -- (Ch5);
  \draw[ha,gray!60] (Ch1) -- (Cai);
  \draw[ha,gray!60] (Ch3) -- (Cai);
  \draw[ha] (Ch5) -- (Cai);
\end{scope}
\node[caplab,below=4pt of C] {(C) Human--human-around-AI \\ \footnotesize\mdseries longitudinal, multi-channel};

\end{tikzpicture}
  \caption{Three triadic configurations. The arrow into the AI agent in each
  panel is the locus of current public corpora; the dashed arrows between
  humans are the locus of the missing training signal.}
  \label{fig:triadic}
\end{figure}

All three share a property: the substantive engineering signal is in the
human--human portion, with the human--AI portion as downstream consumer or
executor. Figure~\ref{fig:triadic} sketches the three configurations.

\subsection{Properties of triadic data}

Triadic data differs from existing public corpora along three axes that
matter for both methodology and licensing.

\textbf{Synchrony.} Triadic data is captured in real time, with all
modalities aligned to a single timeline at sub-second resolution: audio per
participant, screen content, IDE state, structured action streams, terminal
output. Post-hoc transcript-only data loses the screen and IDE state where
most engineering decisions are anchored. The alignment infrastructure is
mature in egocentric vision research~\citep{ego4d2022,epicKitchens2018}, and
the same instrumentation now deployed in consumer AR/VR and industrial
wearables transfers directly to engineering settings.

\textbf{Consent and sanitization.} Triadic data is captured from real
engineering work, often involving proprietary or sensitive material.
Sanitization must occur at capture time, not post-hoc. We propose a layered
approach: automated named-entity recognition for hostnames, function names,
and customer references, with redaction overlays; automated audio
diarization and silencing of identifiable customer names; human-in-the-loop
verification on a calibrated sample. Methods from medical-imaging
deidentification and protected-speech ASR provide direct precedent.
Crucially, simulated companies---fictional companies populated by real senior
contributors---are the configuration where consent and sanitization are most
tractable, because the proprietary problem is upstream-engineered out.

\textbf{Ambiguity tolerance.} Triadic data does not have a single ground
truth. Two senior engineers may produce different correct designs for the
same problem. Methods from affective computing, where ground truth is
fundamentally uncertain, provide the relevant prior art. We adopt the
principle that rubrics should target inter-rater agreement at $\kappa \ge
0.75$ for \emph{categories of events} rather than for specific labels, and
that disagreement should be preserved in the released dataset rather than
majority-voted away. This is a methodological shift the SWE community is
unaccustomed to but the perception community has practiced for decades.

\subsection{What the triadic frame is not}

The triadic reframing is not a rejection of dyadic data; we explicitly
endorse the call to collect human--AI dialogue at scale. Nor is it a
recommendation for surveillance: every methodology described here is
consent-based, applied to volunteer participants, with explicit deletion
rights. Nor is it a claim that triadic context replaces solo benchmarks:
SWE-bench Pro~\citep{sweBenchPro2025}, Terminal-Bench~\citep{terminalBench2026},
SWE-EVO~\citep{sweEVO2025}, and APEX-SWE~\citep{apexSWE2026} remain essential
evaluation substrate. The triadic frame is a claim about training data, not
about evaluation.

\section{Two Complementary Data Products}\label{sec:products}

A position paper that names a substrate without specifying capture is
incomplete. We outline two complementary products that together instantiate
the triadic frame.

\subsection{Long-horizon expert trajectories with stimulated recall}

Senior engineers work through real debugging sessions and feature-implementation
tasks in an instrumented environment---IDE telemetry, screen recording,
terminal capture---followed by a stimulated-recall walkthrough within thirty
minutes, in which the engineer re-watches the session and verbalizes their
reasoning. The hybrid passive-then-recall protocol draws on a forty-year HCI
tradition~\citep{ericsson1993protocol} and produces richer reasoning traces
than concurrent think-aloud, with substantially less behavioral distortion of
the underlying work.

Each trajectory is delivered as a structured object: timestamped IDE events,
screen segments keyed to events, recall transcript with timestamps, final
code diff, and step-level annotations defined at test-execution boundaries.
The result is a reasoning trace with verifiable checkpoints---directly
compatible with supervised fine-tuning, with preference-pair construction
(expert versus baseline-model trajectory on the same task), and with process
reward modeling at step granularity.

The trajectory mix should be deliberately weighted toward the long horizon:
a substantial majority of capture time on tasks of one to four hours and an
explicit minority on multi-hour extended sessions, with shorter-horizon work
as anchor distribution rather than primary signal. Per-trajectory cost is the
binding economic constraint---at realistic senior-engineer rates, four-hour
capture with recall is order-of-magnitude \$1{,}000 per trajectory---but the
per-trajectory eval delta on long-horizon evaluations should be commensurately
larger than for short-horizon data. The empirical question of whether expert
traces yield disproportionate eval uplift, in the spirit of
LIMA's~\citep{zhou2023lima} curated-over-quantity result for instruction
tuning, is the first question the trajectory product is designed to answer.

\subsection{Simulated cross-functional companies}

The expert-trajectory product captures individual senior reasoning. It does
not capture cross-functional friction---the negotiation between engineers,
product managers, designers, and data scientists that produces ambiguous
requirements, contested architectural decisions, and the lived practice of
interpretation. This is what simulated companies are for.

A simulated company is a fictional company---fictional product, fictional
customer base, fictional internal politics---staffed by real senior
contributors in their actual professional roles. Teams of four to six
(engineers, PM, designer, data scientist, occasional security or SRE
specialist) work through one-to-three week deliverables using the standard
tools of distributed engineering: Slack or equivalent for asynchronous
discussion, GitHub for code, design tooling for mockups, video conferencing
for synchronous meetings. All channels are instrumented with API capture; all
participants sign explicit consent; all sensitive personal information is
scrubbed at capture.

The output of a single project is dense and qualitatively different from
anything in current corpora. Hundreds of conversational turns across roles.
Dozens of document revisions tracking how requirements evolve under
negotiation. Design artifacts with version history. A commit graph with
associated code review threads. A decision log linking design conversations
to architectural choices to implementation. Crucially: the \emph{connections}
between channels---the Slack discussion that produced the design doc that
produced the ticket that produced the PR that produced the code review
feedback---are preserved.

The configuration matters. Simulated companies sidestep the proprietary-data
problem that makes Configuration C (real human--human-around-AI work) hardest
to license. Because the company is fictional, IP and confidentiality concerns
are bounded. Because the contributors are real senior practitioners, the
engineering signal is preserved. Because the projects span weeks rather than
hours, the data captures the temporal scales where current agents fail.

The first-order economic question for simulated companies is whether the
per-project cost (substantial---multiple senior contributors for multiple
weeks) yields training signal proportionate to the investment. We propose
that this is empirically tractable at a pilot scale of a small number of
projects, with frontier-lab evaluation of the resulting data the natural
go/no-go gate.

\section{A Four-Tier Evidence Framework for Corpus Quality}\label{sec:quality}

A high-cost corpus is only as useful as its credibility to a fine-tuning
researcher. The two products of Section~\ref{sec:products} are both expensive,
both novel, and both make claims that cannot be verified by reading the
artifact alone. We therefore specify a four-tier evidence framework---adapted
from a prior memo by the author~\citep{kim2026dataquality}---through which a
triadic corpus, or any post-training corpus, must justify its quality
quantitatively. The framework is orthogonal to the triadic claim: it applies
equally to GitHub scrapes, agent self-trajectories, dyadic dialogue, and
triadic capture. Its inclusion here is to make the implicit empirical contract
between data producer and fine-tuning consumer explicit, since this contract
is precisely where most public corpora fail in the after-action review.

Quality operates at three levels (Table~\ref{tab:levels}). Item-level
correctness can be 99\% and the corpus can still be useless; corpus-level
distribution can be exemplary and the data still fails to teach. Each level
has different failure modes and different verification strategies.

\begin{table}[t]
\small
\caption{Three levels at which SWE training-data quality must be evaluated.
Item-level checks are necessary but never sufficient.}
\label{tab:levels}
\centering
\begin{tabular}{@{}lll@{}}
\toprule
Level & Unit of analysis & Example failure mode \\
\midrule
Item     & A single (prompt, response), (issue, patch), or trajectory & Patch contains a hallucinated import \\
Corpus   & The full dataset                                          & 80\% Python-web bugs; misses target distribution \\
Training & How data behaves under actual fine-tuning                 & Teaches format but hurts capability \\
\bottomrule
\end{tabular}
\end{table}

\begin{figure}[t]
  \centering
\begin{tikzpicture}[
  tier/.style={draw=black!60,rounded corners=4pt,minimum width=3.5cm,
               minimum height=1.6cm,align=center,font=\small,inner sep=4pt,fill=#1},
  arr/.style={->,>=Stealth,thick,black!70},
  reaction/.style={font=\scriptsize\itshape,align=center,text=black!55},
  bracket/.style={decorate,decoration={brace,amplitude=4pt,raise=2pt}}
]

\node[tier=blue!8]   (T1) at (0,0)    {\textbf{T1: Mechanical} \\[1pt] compile, AST,\\imports, tests,\\leakage, license};
\node[tier=blue!14]  (T2) at (4.2,0)  {\textbf{T2: Statistical} \\[1pt] distribution,\\ diversity,\\difficulty calibr.};
\node[tier=blue!22]  (T3) at (8.4,0)  {\textbf{T3: Probes} \\[1pt] small-model FT,\\learning curves,\\ablations};
\node[tier=blue!32]  (T4) at (12.6,0) {\textbf{T4: Pre-reg.\ blind} \\[1pt] predict$\to$deliver$\to$\\evaluate blind,\\ measure delta};

\draw[arr] (T1) -- (T2);
\draw[arr] (T2) -- (T3);
\draw[arr] (T3) -- (T4);

\node[reaction,below=4pt of T1] {``Baseline hygiene.''};
\node[reaction,below=4pt of T2] {``I see the shape.''};
\node[reaction,below=4pt of T3] {``You're serious.''};
\node[reaction,below=4pt of T4] {``Now I can reason\\about you.''};

\node[font=\footnotesize,above=2pt of T1.north,xshift=0pt,text=black!60] {necessary};
\node[font=\footnotesize,above=2pt of T4.north,xshift=0pt,text=black!60] {sufficient (with all prior)};

\node[font=\scriptsize,below=1.8cm of T2.south east,xshift=2cm,
      text=black!70,align=center] {Increasing evidence strength $\longrightarrow$ \\
      Item-level $\to$ Corpus-level $\to$ Training-level $\to$ Out-of-sample};

\end{tikzpicture}
  \caption{Four-tier evidence framework. Item-level mechanical checks (T1)
  and statistical corpus characterization (T2) are necessary baselines.
  Probe experiments (T3) and pre-registered blind evaluation (T4) are where
  most public corpora fail---and where credible producers distinguish
  themselves.}
  \label{fig:tiers}
\end{figure}
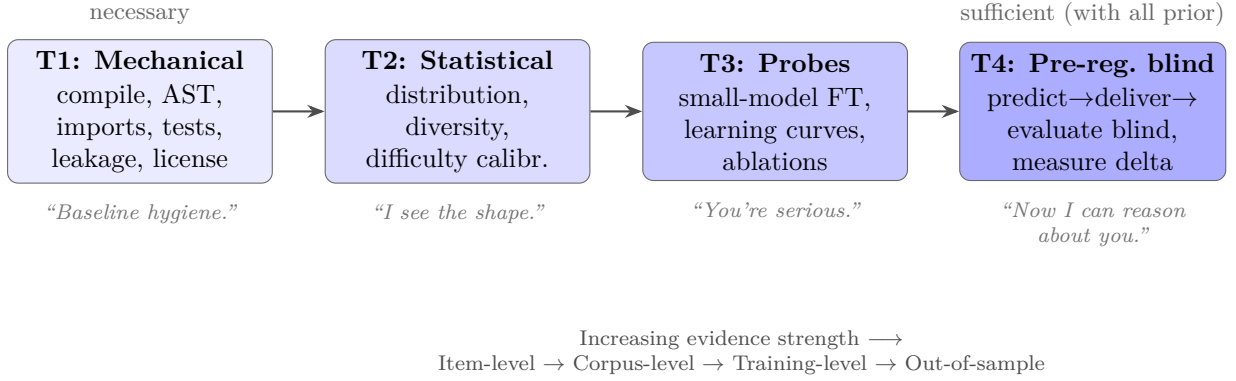

The four tiers below correspond to a progressively stronger evidence package
that a corpus producer can offer a fine-tuning researcher. Each tier is
necessary; none is alone sufficient. Figure~\ref{fig:tiers} sketches the
progression.

\subsection{Tier 1: Mechanical verification}

For each item, automated checks: code compiles; AST parse succeeds; imports
resolve against the commit-time environment; tests pass when applicable; the
patch does not modify test files; the issue text does not leak the solution;
licensing is verified; PII and secrets are scanned. These are mechanical.
Most vendors do the first three and skip the others; the latter two---test
modification and solution leakage---are precisely where SWE-bench+ found the
threefold inflation~\citep{swebenchPlus2024}.

Beyond mechanical checks, semantic verification is harder but available:
adversarial test generation (can we construct a test the patch should also
pass but does not?), differential mutation testing (does a high-quality fix
reject mutations that a band-aid fix accepts?), and timestamp-aware API
resolution (does the patch reference APIs that did not exist at commit time?).
For triadic corpora specifically, semantic checks must also include
\emph{conversational consistency}: are the IDE events, audio transcript, and
post-hoc recall internally coherent at every checkpoint?

\subsection{Tier 2: Statistical corpus characterization}

The corpus is described, not just sampled. Distribution across languages,
frameworks, task types, codebase sizes, and difficulty levels is reported
quantitatively. Diversity is measured as embedding clustering entropy, AST
fingerprint entropy, and pairwise n-gram overlap. Difficulty is calibrated
against a baseline model: items where the baseline's success rate is in the
range $[0.2, 0.7]$ are at the model's frontier and carry the most learning
signal; items the baseline solves trivially carry none. Contamination is
audited explicitly against standard benchmarks (SWE-bench, HumanEval,
LiveCodeBench) using MinHash, suffix-array exact match, and semantic
similarity for paraphrase contamination.

For triadic corpora, the corpus-level dimensions extend naturally: distribution
across triadic configurations (A/B/C), distribution across cross-functional
roles (engineer/PM/designer/data scientist), distribution of session
durations, and distribution across the five sub-categories of implicit
knowledge invocation defined in Section~\ref{sec:directions}.

\subsection{Tier 3: Probe experiments}

The corpus producer fine-tunes a small open model (e.g., Qwen-7B,
Llama-3-8B) on the corpus and characterizes what it does. Four probes:
(P1) small-model sanity check on the claimed target capability, adjacent
capabilities, and unrelated capabilities (regression check); (P2) learning-curve
analysis at 1K, 5K, 10K, 50K items, looking for monotonic smooth improvement
versus rapid plateau versus contamination-induced spikes; (P3) within-corpus
ablation, splitting on language or task type and training on each split
separately; (P4) adversarial evaluation under prompt paraphrase, context
shuffle, and translation, distinguishing real capability from formatting
artifact.

Probe experiments are where most vendor reports stop being marketing and
start being science. A producer who has done these is one who has already
caught issues a researcher would otherwise discover six weeks into
integration.

\subsection{Tier 4: Pre-registered blind evaluation}

Before delivering the data, the producer pre-registers with the researcher:
the evaluation methodology, the expected uplift on the researcher's target
eval, the expected non-regression on other capabilities, and the conditions
under which the data is deemed successful. The data is delivered with held-out
splits the researcher controls. Experiments are run blind. Actual results are
compared against the pre-registered predictions.

The asymmetry is deliberate. A producer whose predictions match reality has
demonstrated they understand their own data. A producer whose predictions
miss has produced honest signal that will recalibrate them. Both outcomes
are scientifically valuable; the alternative---data delivery without
prediction---is not.

The single empirical bar that distinguishes a frontier-lab-caliber data
producer from a vendor: \emph{can you tell a fine-tuning researcher, before
they train anything, what their eval scores will be after training on your
data, to within 1--2 percentage points?} If yes, the producer has internalized
the data. If not, they have only catalogued it.

\section{Research Directions This Substrate Enables}\label{sec:directions}

We outline four research directions enabled by triadic-context data. These
are tractable handles, not exhaustive.

\textbf{Implicit-knowledge invocation modeling.} Senior engineers continually
invoke context that is not in the codebase: ``wait, region X is on the older
Kubernetes version,'' ``we tried that two years ago and it didn't work
because of throughput,'' ``that's the wrong abstraction.'' A model trained on
annotated triadic data could learn to detect these moments in real time,
propose to elicit the implicit knowledge, and update its working context
accordingly. This is a sequence-labeling problem with multimodal inputs, with
direct precedents in dialogue-act tagging and emotion-event detection.

\textbf{Drift-aware agent training.} Solo-trajectory training never teaches
that system state changes mid-task. Triadic data captured longitudinally over
hours and days contains many examples of state changes---a teammate scales a
cluster, a design doc is revised, an upstream service deprecates an API.
Agents trained on this data can learn to detect and re-plan around drift.
Drift-injected training environments are the natural verifiable-reward
complement.

\textbf{Cross-functional coordination learning.} The capability-gap synthesis
of Section~\ref{sec:insufficient} places ambiguous-requirement handling and
architectural reasoning among the top open problems. These are inherently
cross-functional. Simulated-company data, captured in Configuration~C, is the
supervised substrate for learning how senior engineers translate vague PM
input into bounded technical specifications, how designers' constraints
propagate through implementation, and how data scientists' analytical
questions become engineering work. The training problem here is
hierarchical---per-channel summarization, cross-channel decision tracing,
project-level outcome prediction---with direct precedents in long-horizon
planning and meeting-summarization research.

\textbf{Disagreement-grounded reward modeling.} The APEX-SWE epistemic-discipline
gap~\citep{apexSWE2026} suggests that reward models trained on non-expert
preferences over routine problems will not generalize to expert-level work.
We propose that triadic data naturally generates \emph{expert disagreement} on
expert-level problems---two senior pairs solving the same simulated-company
task in different but defensible ways---which provides a direct training
substrate for reward models that learn to detect ``locally plausible but
globally naive'' solutions. Existing preference datasets typically contain
non-expert preferences on routine problems; the disagreement-set methodology
inverts this.

\section{What This Approach Deprioritizes---and What Would Invalidate It}\label{sec:deprioritize}

A position paper that does not explicitly name what it argues against is
unfalsifiable. We name both the alternatives we argue should be deprioritized
and the empirical results that would invalidate the position.

\subsection{Deprioritizations}

\textbf{Competitive-programming-style data.} HumanEval is saturated and
likely contaminated. Competitive programming is single-file, well-specified,
deterministically tested---the opposite of real engineering. We argue this
data should drop from a substantial fraction of code-training mix to under
5\%, replaced by long-horizon and cross-functional capture.

\textbf{Single-file isolated-function generation.} Real production code
imports from dozens of modules, maintains complex state, and interacts with
infrastructure. The SWE-bench+ inflation result (12.47\% $\to$ 3.97\% with
leakage filtered) suggests ``realistic'' benchmarks built on isolated-function
reasoning can be gamed.

\textbf{Python-only annotation pipelines.} The performance gap between Python
and non-Python SWE-bench is a function of annotator-pool composition as much
as model behavior. The remediation is language-specialist annotators, not
Python generalists working in unfamiliar territory.

\textbf{Tutorial-quality and toy data.} GitClear's churn analysis is
consistent with models that have learned tutorial idioms. Production
codebases---large, messy, historically encrusted---should be weighted heavily
over tutorial code in any curated mix.

\subsection{What would invalidate this position}

The position rests on three empirical claims, each testable.

\emph{First:} that triadic-data-trained models outperform dyadic-data-trained
models on long-horizon, drift-prone, multi-team tasks. If carefully matched
experiments show no differential, the position is wrong. We do not yet have
these experiments; the proposed datasets are the experimental apparatus that
would run them.

\emph{Second:} that simulated-company data carries training signal proportionate
to its capture cost. If pilot-scale simulated companies do not yield measurable
evaluation uplift on long-horizon and cross-functional tasks, the
simulated-company instantiation is wrong even if the triadic frame is right.
The first-order pilot is small (single-digit number of projects) and the
go/no-go gate is frontier-lab evaluation.

\emph{Third:} that the agentic flywheel---synthetic data plus verifiable
rewards plus larger models---does not absorb the long-horizon gap on its own.
If self-play with verifiable rewards continues to produce gains at the rate of
the past twelve months, the marginal value of expensive expert capture
declines. The DeepSeek-R1~\citep{deepseekR12025} result and the APEX-SWE
plateau~\citep{apexSWE2026} both have to be re-read in eighteen months.

We commit, on our side, to publishing the empirical results of pilot-scale
capture under permissive license regardless of outcome.

\section{Future Directions}\label{sec:future}

Three directions deserve naming as natural extensions of triadic methodology
beyond the position's central claim.

\textbf{Visual and perception-native SWE.} SWE-bench
Multimodal~\citep{swebenchMultimodal2024} places frontier systems at 12\%
resolution on visual JavaScript bugs. The triadic-data methodology extends
naturally to visual SWE: pair sessions on frontend bugs,
design-mockup-to-implementation traces, accessibility audits, design-review
conversations grounded in screen content. The capture infrastructure
(synchronized screen plus IDE plus audio) is the same. The annotation
taxonomy needs extension, but the precedents are mature in egocentric vision
and human-activity recognition. This is a high-conviction direction for any
group with deep perception expertise: pure-LLM researchers are bolting vision
onto text-native architectures; the field is short on people who think
natively in spatial and visual grounding.

\textbf{Security-focused expert capture.}
\citet{perryInsecure2023}'s confidence-competence inversion---AI-assisted
developers producing less secure code while believing the opposite---is
uniquely dangerous because it scales harm rather than productivity. Triadic
capture of security engineers and penetration testers, in pair sessions on
real vulnerability investigations, produces training signal that no scrape
of GitHub will. The annotator pool here is small (security specialists, not
coding generalists) and the consent protocols are stricter (real-CVE work
requires careful de-identification), but the per-trajectory eval delta is
correspondingly large.

\textbf{Formal verification as RL reward.} For safety-critical
domains---embedded systems, medical devices, aerospace, financial
infrastructure---probabilistic correctness is insufficient. Formal
verification tools have reached the maturity to serve as automated reward
signals in reinforcement learning, complementing test-suite pass/fail with
stronger correctness guarantees. The training substrate here is not human
capture but tooling integration; the data agenda is to identify the corpus
of formally-specifiable problems where this is operationally feasible.
Triadic data has a contribution here too: senior practitioners' reasoning
about \emph{what to formally verify and what to leave as test-suite-checked}
is itself a training signal worth capturing.

\section{Conclusion}

The next substantial gains in frontier SWE agents will come from training
data that captures how senior engineers handle context, not from more solo
trajectories or harder benchmarks. Existing data approaches---GitHub scrapes,
agent self-trajectories, dyadic human--AI dialogue---are necessary but
insufficient. The deeper substrate is triadic: the human--human conversations
where context is formed, the human--AI sessions where it is consumed, and
the cross-functional work that surrounds both. Two complementary products
instantiate this substrate---long-horizon expert trajectories with
stimulated-recall capture, and simulated cross-functional companies---and a
four-tier evidence framework specifies how either product justifies itself
to a fine-tuning researcher. The capture methodology is already mature in
adjacent fields. The legal and ethical infrastructure is operational. The
empirical case is testable in a 12-to-18-month timeframe. The field's
near-term research agenda should include this work explicitly, and the case
for it should be made before frontier labs begin asking for the data---which
we expect within twelve months regardless of whether the literature has
caught up.

\bibliography{refs}

\end{document}